\documentclass[article,letterpaper,floatfix,showpacs,amsmath]{revtex4}
\newlength{\picwidth}
\usepackage[dvips]{graphicx}
\usepackage{dcolumn}

\begin{document}
\def\eps{\epsilon_s}
\def\leftangle{\left\langle}
\def\rightangle{\right\rangle}
\def\be{\begin{equation}}
\def\ee{\end{equation}}
\def\rhoann{\rho_{\rm ann}}
\def\rhobh{\rho_{\rm BH}}
\def\rhonead{\rho_{1,{\rm ad}}}
\def\ba{\begin{eqnarray}}
\def\ea{\end{eqnarray}}
\def\vvec{{\mbox{\boldmath $v$}}}
\def\grad{{\mbox{\boldmath $\nabla$}}}
\def\dotprod{{\mbox{\boldmath $\cdot$}}}
\def\Mann{M_{\rm ann}}
\def\Gamann{\Gamma_{\rm ann}}
\def\sigv{\langle\sigma v\rangle_{\rm ann}}
\def\vrms{\langle v^2\rangle}
\def\rann{R_{\rm ann}}
\def\gvec{{\mbox{\boldmath $g$}}}
\def\rvec{{\mbox{\boldmath $r$}}}
\def\msun{M_\odot}
\def\lcdm{\Lambda{\rm CDM}}
\def\omcdm{\Omega_{\rm CDM}h^2}
\def\gstar{g_\star}
\def\gstars{g_{\star S}}
\def\gstarsk{g_{\star SK}}
\def\rza{R_{\rm ZA}}
\def\gev{{\rm GeV}}
\def\cm{{\rm cm}}
\def\kcdm{K_{\rm CDM}}
\def\Mhat{\hat M}
\def\Rhat{\hat R}
\def\vff{v_{\rm ff}}
\def\Dm{\Delta M}
\def\mshell{m_s}
\def\rshellt{r(\mshell,t)}
\def\tzm{t_0(\mshell)}
\def\rhomt{\rho(\mshell,t)}
\def\vmt{v(\mshell,t)}
\def\phimt{\phi(\mshell,t)}
\def\vann{v_{\rm ann}}

\title{Spherical Gravitational Collapse of Annihilating
Dark Matter and the Minimum Mass of CDM Black Holes}

\author{R. Ali Vanderveld}
\affiliation{Center for Radiophysics and Space Research, Cornell University, Ithaca NY 14853}
\author{Ira Wasserman}
\affiliation{Center for Radiophysics and Space Research, Cornell University, Ithaca NY 14853}

\begin{abstract}

Spherical gravitational collapse of a cold gas of annihilating particles
involves a competition between two density dependent rates:
the free-fall rate $\propto\sqrt{\rho}$ and the (s-wave) annihilation
rate $\propto\rho$. Thus, there is a critical density $\rhoann$
above which annihilation proceeds faster than free fall. Gravitational
collapse of a cloud of (initial) mass $M$ to a black hole is only
possible if $\rhobh\equiv 3M/4\pi(2GM)^3=3/32\pi G^3M^2\lesssim\rhoann$,
which implies that $M\gtrsim\Mann\equiv (3/32\pi G^3\rhoann)^{1/2}$.
We compute the collapse and annihilation of cold, spherical gas clouds with
either a homogeneous density profile or a specific radially-dependent
profile that would lead to self-similar collapse if there were no
annihilation. These calculations verify that there exists a lower bound to the range of possible black hole masses that can form in cold dark matter (CDM) collapse.
For a particle mass $m$ and freeze-out temperature $T_f=m/x_f$,
the minimum black hole mass is $\Mann\approx 10^{10}\msun
\times\left(x_f\sqrt{g_\star}/100\omcdm g_{\star S}m({\rm Gev})
\right)$,
where $g_{\star S}$ and $g_\star$ are degeneracy factors.  Moreover, the formation of a black hole is accompanied by the annihilation of about $M_{ann}$ of the original mass.  Most of the annihilated mass is released in a burst lasting a time $\sim GM_{BH}$, where $M_{BH}$ is the black hole mass when collapse first occurs.  A total annihilation luminosity $\sim 10^{59}~{\rm erg~s^{-1}}$ could easily be reached, lasting a time $\sim 100$ seconds.  The absence of astronomical observations of such spectacular events suggests either: (i) the branching ratio for CDM annihilation to $e^{+}e^{-}$ pairs or quarks $\lesssim 10^{-10}$, while the branching ratio to $\nu{\overline{\nu}}$ is $\lesssim 10^{-5}$; or (ii) CDM is not made of annihilating particles, but may be in some non-annihilating form, such as axions; or (iii) CDM black holes never form.
\end{abstract}
\maketitle

\section{Introduction}

Non-baryonic matter dominates the matter content of the Universe
today, with a density parameter $\omcdm\approx 0.15$; 
the rest of the energy density is primarily in the form 
of dark energy \cite{WMAP1, Status}. In the favored
cold dark matter (CDM) scenario based on weakly interacting massive particles (WIMPS), the dark matter is in the form
of a gas of particles that is very cold today, with an abundance
that was lowered substantially by annihilation when the temperature
of the Universe was $\gtrsim m/x_f$, for a CDM particle mass $m$
freezing out at temperature $T_f=m/x_f\sim 0.1m$ \cite{LowerBound, Conseq, Candidate}. The $\lcdm$ picture provides acceptable fits to the 
temperature and E-polarization fluctuations of the cosmic microwave
background (CMB) radiation \cite{WMAP2, WMAP3}, as well as to the large scale
structure of the Universe \cite{SDSS1}. On smaller scales, CDM
halos seem to form cuspy cores \cite{Cusps, NFW}; dark matter annihilation
has been proposed as one means of smoothing out the cusps
\cite{Cold, Ann1}. CDM clumps are found to form in large scale
structure simulations and to agglomerate to form halos \cite{Small};
annihilation within these clumps would lead to an enhanced
flux of cosmic ray products at Earth \cite{GC, Ann2}.

In this paper, we investigate a related question: How does annihilation alter 
spherically symmetric collapse of CDM? We
consider, for the most part, the collapse of CDM clouds with
zero or very small internal energy. In the absence of annihilation,
the inevitable result would be complete gravitational collapse
to a black hole. 

However, to form a black hole, a collapsing sphere of mass $M$
must reach a mass density of order $\rhobh=3M/4\pi(2GM)^3=
3/32\pi G^3M^2$. When (s-wave) annihilation is included, the 
rate of CDM annihilation at a density $\rho$ is
\be
\Gamann={\rho\sigv\over m}~,
\ee
where $\sigv$ is the annihilation rate coefficient and $m$ is
the CDM particle mass. For comparison, the free fall rate
at high densities is approximately
\be
H=\sqrt{8\pi G\rho\over 3}~;
\ee
then $\Gamann=H$ at a density
\be
\rhoann={8\pi Gm^2\over 3\left(\sigv\right)^2}~.
\label{rhoann}
\ee
For $\rho\lesssim\rhoann$, free fall proceeds faster than annihilation,
and vice-versa. Therefore, black hole formation is only possible
for a mass $M$ provided that $\rhobh\lesssim\rhoann$ or
\be
M\gtrsim{3\sigv\over 16\pi G^2m}\equiv\Mann~.
\label{mann}
\ee
The annihilation of CDM particles in the early Universe leads to 
a residual density parameter $\omcdm$ provided that
\be
\sigv\approx 8.6\times 10^{-9}~{\rm GeV^{-2}}\left({x_f\over
100\omcdm \gstars/\sqrt{\gstar}}\right)~,
\ee
where $\gstars$ and $\gstar$ are degeneracy factors for the
entropy and mass density when the temperature of the Universe
is $\sim T_f$, and
\be
x_f\approx\ln C-{1\over 2}\ln(\ln C)~,
\ee
where $C=0.038(g/\sqrt{\gstar})m\sigv/\sqrt{G}$, with $g$ being
the degeneracy factor for the CDM particles alone \cite{KT}. Substituting for $\sigv$ in equation (\ref{mann}),
we find
\be
\Mann\approx {10^{10}\msun\over m({\rm GeV})}\left({x_f\over
100\omcdm \gstars/\sqrt{\gstar}}\right)
\approx {10^{10}\msun\over m({\rm GeV})}
\left({x_f\over 15\gstars/\sqrt{\gstar}}\right)~,
\ee
where $m=m({\rm GeV})~{\rm GeV}$; furthermore, in our last result, we assume that $\omcdm\approx0.15$, as has been determined
from analyses of the CMB and of large scale structure. Our estimate of
$\Mann$ depends on the CDM particle mass in several ways, since 
$\Mann\propto x_f\sqrt{\gstar}/m\gstars$. In addition to the 
explicit inverse proportionality with $m$, there is a logarithmic
dependence from $x_f$ and a piecewise flat dependence via 
$\gstars$ and $\gstar$: for example, if $m=50~{\rm GeV}$, we would
have enough particle species for $\gstar\approx\gstars\approx 50$,
and $\Mann\approx 4\times 10^7\msun$ (taking $x_f\approx 20$ for
this case). Plausibly, the value of the minimum black hole mass
for CDM is comparable to the mass range deduced for black holes
in the centers of galaxies \cite{BH}. It is remarkable (although
perhaps fortuitous) that the value of $\Mann$ is even remotely
close to an astronomically significant value.

Spherical collapse is, of course, highly idealized. More realistically,
we would expect a collapsing cloud to have substantial substructure
as well as large scale shear flows induced by the tidal field of
nearby fluctuations. However, we would still expect that CDM clouds
that are, however improbably, nearly spherical and smooth would have
the {\it best chance} to collapse to black holes. Thus,
our lower bound on the CDM black hole mass ought to be robust.
Moreover, shear enhances the collapse of a cloud (as is evident
from the Raychauduri equation, e.g. \cite{Peebles}) and anisotropic
collapse of a cloud with internal fluctuations could trigger
substantial fragmentation (e.g. \cite{LoebRasio, VMSV} and refs. therein). It is possible that the calculations 
we present apply to late time collapse within a larger cloud that 
has fragmented considerably.

Section II presents calculations of the collapse of homogenous
CDM spheres. Section III computes the collapse of a specific model with
radial density dependence. In each case, we quantify when black
holes may form, and when the collapsing mass will instead annihilate
away. The detailed calculations are Newtonian, and assume
instantaneous escape of the (relativistic) products
of CDM annihilation. We expect that our main results
for the critical mass for black hole formation, and
for the mass-energy and luminosity released by 
annihilation, to remain valid within factors 
$\sim 1$ in a general relativity calculation that
also accounts for the opacity of the infalling CDM to
the annihilation products.  In Section IV, we discuss our results, including a brief account of what might happen in the collapse of CDM with predominantly p-wave
annihilation.

\section{The homogeneous case}

\subsection{Equations of motion}

Let us consider a uniform sphere of matter with time-dependent radius and mass, called $R(t)$ and $M(t)$, respectively; hence, the density at a time $t$ is $\rho(t)=3M(t)/4\pi R^{3}(t)$. We assume that the homogeneous sphere collapses under its own
self gravity, but we also include pressure forces. To maintain
homogeneity, we assume that
\be
P(r,t)=P_0(t)\left[1-{r^2\over R^2(t)}\right]
\ee
at time $t$ and Eulerian radius $r$; we also assume that
$P_0(t)=K\rho^\gamma(t)$, where $K$ is independent of time.
The Euler equation then becomes 
\be
\ddot{R}=\frac{2K(3M/4\pi)^{\gamma-1}}{R^{3\gamma-2}}-\frac{GM}{R^{2}}~.
\label{eq:eom1}
\ee
If $\gamma>4/3$, then there exists a radius $R_{ZA}$ at which $\ddot{R}=0$.  For a nonrelativistic gas we expect to have $\gamma=5/3$, so in this case we have the zero acceleration radius
\be
R_{ZA}=\frac{2K}{GM^{1/3}}\left(\frac{3}{4\pi}\right)^{2/3}~.
\ee
This is a minimum radius at which the pressure force becomes as great as the gravitational force such that the collapse will ``bounce'' back outward.  In the absence of annihilation, we also have the conserved energy
\be
E=\frac{1}{2}\dot{R}^{2}-\frac{GM}{R}+\left(\frac{3M}{4\pi}\right)^{2/3}\frac{K}{R^{2}}~.
\ee
From this we see that the cloud will bounce at a radius
$R={1\over 2}\rza$ in the absence of annihilation.

We note that there may be three possible physical sources 
of ``pressure'': (i) internal motions associated with small
scale density inhomogeneities within a cloud; (ii) residual
but small entropy of the CDM particles; (iii) trapped
annihilation products. 

Of these three sources, the first
is potentially the most important, at least on large scales, but,
as discussed in the introduction, we are focusing on fragments
that have the best chance for collapse. Thus, we assume that
our collapsing cloud is smooth; the larger scale motions 
associated with inhomogeneities of the parent system are assumed
to be in the form of orbital motion of the subclumps
into which it has fragmented by the time our smooth cloud
collapses to very high density.

The residual entropy of the CDM is small
and depends on the scattering cross section of thermalized
particles in the early Universe with the CDM. Assuming a
cross-section $f_t\sigv(T/m)^2$, the rate of energy transfer
from thermal particles to CDM is $\dot U_t=\zeta(3)g_tf_t\sigv
T^7/\pi^2m^3$ per CDM particle,
where $g_t$ is the spin weight of relativistic
thermal particles that scatter from the CDM and $f_t$ is a numerical factor of order one. The CDM therefore remains in ``kinetic'' thermal
contact until $\Gamma_t=2\dot U_t/3T=H=(8\pi^3\gstar GT^4/90)^{1/2}$,
which implies kinetic freeze-out at \cite{Conseq}
\be
T_K\approx{1.9\gstar^{1/8}G^{1/8}m^{3/4}\over g_t^{1/4}f_t^{1/4}
\left(\sigv\right)^{1/4}}
\approx 3.4\times 10^{-3}~{\rm GeV}~{\gstars^{1/4}[m(\gev)]^{3/4}
\over g_t^{1/4}f_t^{1/4}}\left({100\omcdm\over x_f}\right)^{1/4}~;
\ee
the value of $K$ associated with this residual thermal
energy is then
\ba
\kcdm&=&{T_K\over m[\rho(T_K)]^{2/3}}\nonumber\\
&\approx& 4.9\times 10^8~\gev^{-8/3}
{g_t^{1/4}f_t^{1/4}\over [m(\gev)]^{7/4}\gstars^{1/4}
\gstarsk^{2/3}}\left({x_f\over 100\omcdm}\right)^{1/4}
\left({0.15\over\omcdm}\right)^{2/3},
\ea
where $\gstarsk$ is the degeneracy factor for the 
entropy at temperature $T_K$.
To put this in context, the CDM phase space density is 
$\sim\kcdm^{-3/2}m^{-4}\sim 10^{-13}[m(\gev)]^{-11/8}$.
The CDM phase space density and, therefore, $\kcdm$ are conserved as long as
the CDM particles are collisionless.

The third source of pressure would be from trapped products of
the annihilating dark matter. For the most part, we shall assume
that these annihilation products escape without being trapped,
and this further assumption will be discussed below. (We shall also
re-examine this assumption semi-quantitatively in our brief
discussion of the collapse of CDM that annihilates predominantly
via p-wave interactions.) Any products that are trapped should
be extremely relativistic and therefore would form a $\gamma=4/3$
gas. At fixed $K$, a $\gamma=4/3$ gas either stops the collapse at all
times, or not at all. 

Annihilation implies a mass loss rate
\be
\dot{M}=-\frac{3\sigv M^{2}}{4\pi mR^{3}}~.
\label{eq:eom2}
\ee
Here, we assume that the annihilation products escape the 
collapsing cloud completely. This and equation (\ref{eq:eom1}) give us our system of coupled differential equations which we must solve.  These look much simpler once nondimensionalized with a change to new mass, radius and time units.  Let the mass unit be the initial mass $M_{0}$ such that the new mass variable is $\hat{M}(t)=M(t)/M_{0}$.  Then introduce a radius unit $R_{u}$ and a time unit $\sqrt{R_{u}^{3}/GM_{0}}$ for the new variables $\hat{R}$ and $\tau$, respectively.  Then the above equations become
\be
\frac{d^{2}\hat{R}}{d\tau^{2}}=\frac{R_{ZA}\hat{M}^{2/3}}{R_{u}\hat{R}^{3}}-\frac{\hat{M}}{\hat{R}^{2}}
\ee
\be
\frac{d\hat{M}}{d\tau}=-\frac{3\sigv}{4\pi mG}\sqrt{\frac{GM_{0}}{R_{u}^{3}}}\frac{\hat{M}^{2}}{\hat{R}^{3}}\equiv -\left(\frac{R_{ann}}{R_{u}}\right)^{3/2}\frac{\hat{M}^{2}}{\hat{R}^{3}}~;
\ee
here the characteristic radius at which annihilation becomes
important is defined to be
\ba
\rann&\equiv&{3\sigv\over 4^{4/3}\pi Gm}\left({M_0\over\Mann}
\right)^{1/3}\nonumber\\
&=&{3.8\times 10^{15}~\cm\over m(\gev)}
\left({x_f\over 100\omcdm g_{\star S}/\sqrt{g_\star}}\right)
\left({M_0\over\Mann}\right)^{1/3}~.
\ea
We choose $R_u=\sqrt{\rann^2+\rza^2}$; letting $\kappa=\rza/R_u$
and $\rann=R_u\sqrt{1-\kappa^2}$, the equations describing
the collapse become
\ba
{d^2\Rhat\over d\tau^2}&=&{\kappa\Mhat^{2/3}\over\Rhat^3}-{\Mhat\over
\Rhat^2}\nonumber\\
{d\Mhat\over d\tau}&=&-{(1-\kappa^2)^{3/4}\Mhat^2\over\Rhat^3}~.
\label{cloudev}
\ea
The characteristic timescale for the cloud's evolution is
\ba
t_u&=&\sqrt{R_u^3\over GM_0}={3\sigv\over 4\pi Gm}\left({R_u\over
\rann}\right)^{3/2}\nonumber\\
&\approx&{2.0\times 10^5~{\rm sec}\over m({\rm Gev})(1-\kappa^2)^{3/4}}
\left({x_f\over 100\omcdm \gstars/\sqrt{\gstar}}\right)~.
\ea
For cases where pressure forces are unimportant, $\kappa\to 0$;
for $K=\kcdm$
\be
{\rza\over\rann}={\kappa\over\sqrt{1-\kappa^2}}
\approx{1.3\times 10^{-5}g_t^{1/4}f_t^{1/4}\gstars^{13/12}\over
\gstarsk^{2/3}\gstar^{2/3}[m(\gev)]^{5/12}}
\left({100\omcdm\over x_f}\right)^{13/12}
\left({0.15\over\omcdm}\right)^{2/3}\left(\frac{M_{ann}}{M_{0}}\right)^{2/3}~,
\ee
which is indeed very small.

The solution of equations (\ref{cloudev}) depends on the value of
$\kappa$ and the initial conditions. Well before annihilation
becomes important, $\Mhat$ remains constant, and
\be
\left(\frac{d\hat{R}}{d\tau}\right)^{2}=-\frac{\kappa}{\hat{R}^{2}}+\frac{2}{\hat{R}}+E~.
\ee
For $E=0$, we can integrate to find
\be
\sqrt{2}|\tau-\tau_{bounce}|=\frac{2}{3}\left(\hat{R}-\frac{\kappa}{2}\right)^{3/2}+\kappa\left(\hat{R}-\frac{\kappa}{2}\right)^{1/2}~,
\label{timesol}
\ee
where the bounce happens when $\tau=\tau_{bounce}$.  We use equation (\ref{timesol}) to estimate the initial time for our numerical solutions.

As for the initial conditions, $\hat{M}(\tau)=1$ at the start and $\hat{R}(\tau)$ should ideally fall from a very large distance, where $d\hat{R}/d\tau\approx 0$.  However, since annihilation occurs only at small radii $\hat{R}\lesssim 1$, we can initialize numerical calculations to excellent accuracy by setting $\hat{M}_{0}=1$ as long as the initial value $\hat{R}_{0}$ is sufficiently large.  We can then find the corresponding initial infall velocity by assuming previously negligible annihilation; in this limit, we can use the above energy conservation equation to find
\be
\left(\frac{d\hat{R}}{d\tau}\right)_{0}=-\sqrt{\frac{2}{\hat{R}_{0}}-\frac{\kappa}{\hat{R}_{0}^{2}}}~,
\ee
which will be used in the numerical work described below.

\subsection{Numerical calculation}

\begin{figure}[h]
\includegraphics[width=\picwidth,height =9cm ]{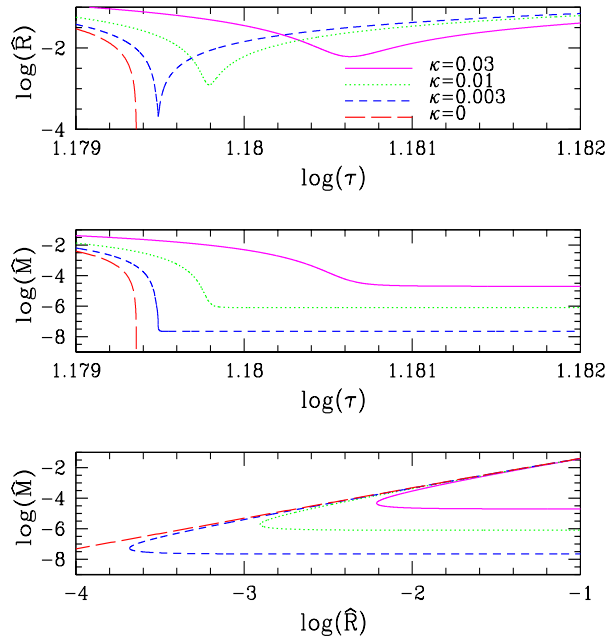}
\caption{The top two plots show $\log(\hat{R})$ and $\log(\hat{M})$ versus $\log(\tau)$, for a time interval near the bounce and for $\kappa=0.03, 0.01,0.003$ and $0$.  The bottom plot is $\log(\hat{M})$ versus $\log(\hat{R})$ for the same four values of $\kappa$.  The initial radius was $\hat{R}_{0}=10$.} 
\label{uniform1} 
\end{figure}

Equations (\ref{cloudev}) were solved using a fourth-order Runge-Kutta routine \cite{NR}.  The initial conditions for the mass, radius and velocity were set as discussed above; $\hat{R}_{0}=10$ turned out to be sufficiently large.  The results, for different $\kappa$ values, are displayed in Figure~\ref{uniform1}.  Note how the minimum $\hat{R}$ and the final $\hat{M}$ values both appear to scale with $\kappa$.  Also, we see that the $\kappa=0$ case yields a power-law relationship between $\hat{M}$ and $\hat{R}$ for small $\hat{M}$ and $\hat{R}$.  We can understand these relationships analytically by looking at the small $\kappa$ and small $\hat{M}$ limit of the equations of motion.

Let us change the independent variable in our equations from $\tau$ to $x=1/\hat{M}$ while defining the new dependent variable $y=\hat{R}^{-2}$.  Rearranging the equations of motion yields
\be
\sqrt{y}\frac{d^{2}y}{dx^{2}}=\frac{2}{x}~.
\ee
This form is more useful for finding the behavior of the solution when $\hat{M}\rightarrow 0$, which is equivalent to $x\rightarrow\infty$.  In this case, let us look for a solution of the form
\be
y=Ax+Bx^{a}+...
\label{eq:zero}
\ee
where the solution will only be sensible if the second term is much smaller than the first.  By inserting this into equation (\ref{eq:zero}) we find that
\be
y\simeq Ax-8\sqrt{\frac{x}{A}}+...
\ee
for large $x$; the coefficient A is not determined by this argument.  Note that, by returning to the original variables, this implies that $\hat{M}/\hat{R}^{2}$ approaches a constant as the sphere collapses.  Putting $\kappa$ back into the equations, but still requiring it to be small, we can take the $\hat{R}\rightarrow 0$ limit in order to predict a final mass $\hat{M}_{f}\propto\kappa^{3}$ and a minimum radius $\frac{1}{2}\hat{R}_{ZA}\propto\kappa^{3/2}$.  These scalings agree with the numerical solutions.

The asymptotic scaling $\hat M/\hat R^2\to{\rm constant}$
for an annihilating cloud implies that $dR/dt\to{\rm
constant}$ at late times. Before annihilation becomes dominant,
the cloud collapses in free fall; this phase ends when
the cloud attains a density $\sim\rhoann$. At that time, the
cloud collapses at a rate $dR/dt\approx\vann\equiv
(M_0/\Mann)^{1/3}$; subsequently, the cloud loses mass,
its self-acceleration diminishes, and it continues
to collapse at roughly this speed. In doing so, its
density is determined by the condition that the rates
of annihilation and collapse remain about the same, i.e.
\be
{\rho\sigv\over m}={3M\sigv\over 4\pi R^3m}
\sim{\vann\over R}~,
\ee
which implies
\be
{M\over R^2}\sim {4\pi m\vann\over 3\sigv}~.
\label{afterann}
\ee
Equation (\ref{afterann}) corresponds to $\hat M/\hat R^2
=A\sim 4^{1/3}$.

At this point, we pause to consider the circumstances under which
our assumption that annihilation products escape promptly is
justified. Let us suppose that, when first produced with individual particle energy
$m$, the relativistic annihilation
products interact with the CDM particles with a cross section
$f_s\sigv$, where we expect $f_s\sim 1$. Then the optical depth
of the collapsing cloud to the relativistic annihilation products
is $\tau=f_s\rho\sigv R/m$. The density at which the cloud first
becomes opaque to these products is
\be
\rho_1=\rhoann f_s^{-3/2}\left({\Mann\over M}\right)^{1/2}
=\rhobh f_s^{-3/2}\left({M\over\Mann}\right)^{3/2}~.
\ee
Thus, we see that the cloud remains transparent to the annihilation
products for cloud masses well above and well below $\Mann$.
For cloud masses $\approx\Mann$, the optical depth rises to $\tau
\sim 1$ just as the cloud reaches $\rhoann\sim\rhobh$. For these
cases, the collapse of the cloud could be altered slightly, delaying
the final outcome -- total collapse or total annihilation -- for
a short time, but probably not for longer than the characteristic
free fall or annihilation time.

\subsection{Black hole formation}

\begin{figure}[h]
\includegraphics[width=\picwidth,height =8cm ]{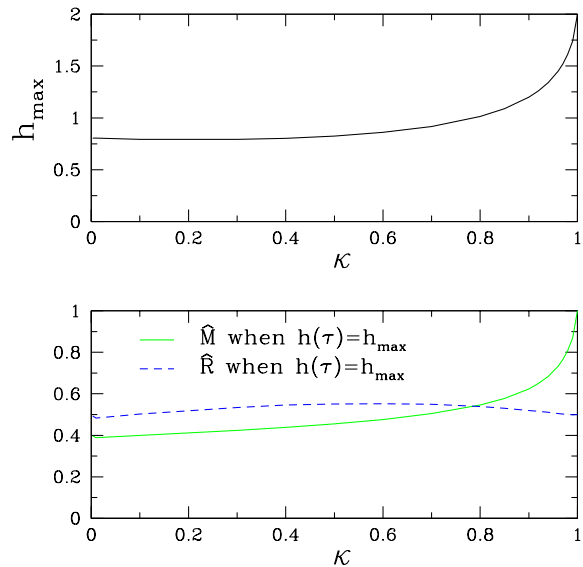}
\caption{The top plot shows $h_{max}$ versus $\kappa$ and the bottom plot shows the values of $\hat{M}$ and $\hat{R}$ at this value of $h$, also versus $\kappa$.  Initial radius $\hat{R}_{0}=10$.} 
\label{uniform2} 
\end{figure}

In order to analyze the black hole formation process, define the dimensionless ``horizon parameter''
\be
h(\tau)=\frac{\hat{M}(\tau)}{\hat{R}(\tau)}=\frac{M(\tau)}{R(\tau)}\frac{R_{u}}{M_{0}}~,
\label{eq:horizon}
\ee 
which increases during the infall phase, reaches a maximum at the bounce and then decreases during the outflow phase.  For $\kappa\neq 0$, this simplifies to
\be
h(\tau)=\frac{2K}{G\kappa}\left(\frac{3}{4\pi M_{0}^{2}}\right)^{2/3}\frac{M(\tau)}{R(\tau)}~,
\ee
and for $\kappa=0$ we get
\be
h(\tau)=\left(\frac{3\sigv}{4\pi m\sqrt{G}M_{0}}\right)^{2/3}\frac{M(\tau)}{R(\tau)}~.
\ee
We can easily find the maximum value of $h(\tau)$ for any particular value of $\kappa$, which ranges from $h_{max}(\kappa=0)\approx 0.8$ to $h_{max}(\kappa=1)=2$.  This is shown in Figure~\ref{uniform2}. In order for a black hole to form, we would require that
\be
\frac{2GM}{R}>1
\label{eq:horizon2}
\ee
and this corresponds to a restriction on $M_{0}$.  For $\kappa\neq 0$, we require
\be
M_{0}\geq\sqrt{\frac{3}{4\pi}} \left(\frac{K}{\kappa G^{2}h_{max}(\kappa)}\right)^{3/4}
\ee
for black hole formation, and similarly for $\kappa=0$ we would require
\be
M_{0}\geq\frac{3\sigv}{4\pi mG^{2}}\left(\frac{1}{2h_{max}(0)}\right)^{3/2}=\frac{\sqrt{2}M_{ann}}{\left[h_{max}(0)\right]^{3/2}}~.
\ee
Therefore there is a definite minimum initial mass $\sim (1-2)M_{ann}$ required in order for the self-gravity of the system to overcome the annihilation and pressure.

\section{A radial density profile}

\subsection{Equations of motion}

A purely collisionless gas cloud, defined to have zero
pressure and to be non-annihilating,
collapsing under its own self-gravity can be visualized
as a set of mass shells in free fall. We can label each shell
by the mass it encloses, $\mshell$, such that its radius at time
$t$ is $\rshellt$. As long as no two shells cross, $\mshell$
remains unchanged and can be used as a comoving coordinate;
this remains true when annihilation is included if we choose
$\mshell$ to be the mass enclosed long before annihilation 
becomes important. For purely collisionless infall,
the motion of an individual mass shell is determined by
its conserved energy $E(\mshell)$ and its ``bang time''
$\tzm$, which is defined by $r(\mshell,\tzm)=0$.

Self-similar solutions for the collapse of a collisionless
cloud have been constructed by Penston \cite{Penston} and Lynden-Bell
and Lemos \cite{LL}, hereafter LL. These models correspond to 
$E(\mshell)=0$ and a power-law $\tzm=t_0+B\mshell^b$; the exact
solution is then
\be
\rshellt=\left({9G\mshell\over 2}\right)^{1/3}
\left(\tzm-t\right)^{2/3}=\left({9G\mshell\over 2}\right)^{1/3}
\left(B\mshell^b+t_0-t\right)^{2/3}~.
\label{rnoann}
\ee
The density and velocity profiles corresponding to this solution are
\ba
\rhomt&=&{1\over 4\pi[\rshellt]^2(\partial\rshellt/\partial m)}
={1\over 6\pi G(Bm^b+t_0-t)[Bm^b(1+2b)+t_0-t]}\nonumber\\
\vmt&=&-\sqrt{2G\mshell\over\rshellt}=-\left({4G\mshell\over 
3(B\mshell^b+t_0-t)}\right)^{1/3}~
\label{rhovnoann}
\ea
where the density profile differs from LL equation (2.9), which has a typo.  Equations (\ref{rnoann}) and (\ref{rhovnoann}) describe a collapse
profile consisting of a {\it shrinking} core of mass 
\be
m_c\sim [(t_0-t)/B]^{1/b}
\ee
and radius 
\be
r_c\sim (18G)^{1/3}B^{-1/3b}(t_0-t)^{(1+2b)/3b}~.
\ee
Within the core, the density is nearly
uniform, $\rho\sim\rho_c=1/6\pi G(t_0-t)^2$, and the
velocity profile is nearly linear, $v(r,t)\sim -r\sqrt
{8\pi G\rho_c}$.  This is surrounded by a halo,
in which the density and velocity profiles are approximately 
powerlaws: 
\ba
\rho&\sim&\rho_c(m_c/\mshell)^{2b}\sim\rho_c(r_c/r)^{6b/(1+2b)}\nonumber\\
v&\sim& v_c(m_c/m)^{(b-1)/3}\sim v_c(r_c/r)^{(b-1)/(1+2b)}~. 
\label{halo}
\ea
In the LL solution, the core density rises to infinity as 
$t\to t_0$, so black hole formation is inevitable in this model.

Our goal is to numerically compute the solution to this problem in the presence of annihilation.  As in the preceding section, we will choose initial conditions such that annihilation will have been negligible previously, making our use of the LL solution a physically meaningful choice.  The above solution has its largest density at the center, so we must specify the central density to be small enough such that it is sufficiently below the critical density specified in equation (\ref{rhoann}).  The central density at the initial time $t=0$ is
\be
\rho(m_{s}=0,t=0)=\frac{1}{6\pi Gt_{0}^{2}}~.
\ee
For our numerical simulations, we require that 
\be
\rho_{ann}=\frac{8\pi\alpha}{3}\rho(m_{s}=0,t=0)~;
\ee
this puts a constraint on the constant $t_{0}$:
\be
t_{0}=\sqrt{\frac{\alpha}{6\pi}}\frac{\sigv}{Gm}~,
\ee
where we choose the constant factor $\alpha$ such that the initial central density is sufficiently small.  We chose $\alpha=3/4\pi\times10^{6}$, which starts the collapse at such an early time that the central density is approximately a million times smaller than $\rho_{ann}$, and therefore the collapse is still in freefall with utterly negligible annihilation.  In this way, we will be justified in initializing our simulation with data from the LL solution.

In order to determine the other constant $B$ for our initial conditions, we can now specify an initial density contrast between the center and the outer edge of the gas cloud; specifying this contrast is also necessary for numerical purposes because the LL solution extends off to infinity.  Let the central density be a factor of $\beta_{i}$ larger than that of the edge at the initial time; this assumption, along with density equation (\ref{rhovnoann}), tells us that we must have
\be
B=\frac{\gamma t_{0}}{M_{0}^{b}}=\sqrt{\frac{\alpha}{6\pi}}\frac{\gamma\sigv}{GmM_{0}^{b}}~,
\ee
where we have once again chosen to call the total initial mass $M_{0}$ and $\gamma$ is the constant
\be
\gamma=\frac{\sqrt{(b+1)^{2}+(2b+1)(\beta_{i}-1)}-(b+1)}{2b+1}~.
\ee
The above conditions give us the constants and initial data necessary for a numerical computation.

Although we are using the initial contrast $\beta_{i}$ for our initial data, the physically important density contrast is what we will call $\beta_{ann}$, which is the contrast when the center begins annihilating.  This is the parameter which will become important when discussing black hole formation for this system.  We can write $\beta_{i}$ as a function of $\beta_{ann}$ and $\alpha$.  First, define the annihilation time $t_{ann}$ to be the time at which the central density would reach the annihilation density in the LL model: 
\be
\rho_{ann}=\frac{1}{6\pi G(t_{0}-t_{ann})^{2}}~.
\label{anntime}
\ee
Then $\beta_{ann}=\rho_{ann}/\rho(M_{0},t_{ann})$, which is
\be
\beta_{ann}=\left(BM_{0}^{b}\sqrt{6\pi G\rho_{ann}}+1\right)\left[\left(1+2b\right)BM_{0}^{b}\sqrt{6\pi G\rho_{ann}}+1\right]
\ee
and we can solve this equation for $BM_{0}^{b}\sqrt{6\pi G\rho_{ann}}$ to get
\be
BM_{0}^{b}\sqrt{6\pi G\rho_{ann}}=-\frac{1+b}{1+2b}+\sqrt{\frac{\beta_{ann}-1}{1+2b}+\left(\frac{1+b}{1+2b}\right)^{2}}~.
\label{B}
\ee
The initial density contrast is
\be
\beta_{i}=\left(\frac{BM_{0}^{b}}{t_{0}}+1\right)\left(\frac{(1+2b)BM_{0}^{b}}{t_{0}}+1\right)
\ee
which can be rewritten as
\be
\beta_{i}=1+2\left(1+b\right)\left(\frac{BM_{0}^{b}\sqrt{6\pi G\rho_{ann}}}{t_{0}\sqrt{6\pi G\rho_{ann}}}\right)+\left(1+2b\right)\left(\frac{BM_{0}^{b}\sqrt{6\pi G\rho_{ann}}}{t_{0}\sqrt{6\pi G\rho_{ann}}}\right)^{2}~;
\ee
plugging in what we know from the equations above, we find
\ba
\beta_{i}=1&+&2\left(1+b\right)\sqrt{\frac{3}{8\pi\alpha}}\left[\sqrt{\frac{\beta_{ann}-1}{1+2b}+\left(\frac{1+b}{1+2b}\right)^{2}}-\frac{1+b}{1+2b}\right]\nonumber\\
&+&\left(1+2b\right)\left(\frac{3}{8\pi\alpha}\right)\left[\sqrt{\frac{\beta_{ann}-1}{1+2b}+\left(\frac{1+b}{1+2b}\right)^{2}}-\frac{1+b}{1+2b}\right]^{2}~.
\ea
We will use $\beta_{i}$ and $\alpha$ in the calculation of initial conditions, but whether or not a black hole forms will be sensitive only to our choice of $\beta_{ann}$.

The addition of annihilation requires that we add the mass loss equation
\be
\frac{\partial\mu(m_{s},t)}{\partial t}=-\frac{\sigv}{m}\rho(m_{s},t)\mu(m_{s},t)
\label{eq:ann}
\ee
where the comoving variable $m_{s}$ now must denote the mass enclosed {\it at the start of the collapse}, while the new dimensionless variable $\mu(m_{s},t)$ is the fraction of mass at a coordinate $m_{s}$ which has not yet annihilated at the later time $t$.  Then the density $\rho$ is simply
\be
\rho(m_{s},t)=\frac{\mu(m_{s},t)}{4\pi r^{2}(m_{s},t)\partial r(m_{s},t)/\partial m_{s}}~.
\label{eq:rho}
\ee
In the presence of annihilation, the collapse equation becomes
\be
\frac{\partial^{2}r(m_{s},t)}{\partial t^2}=-\frac{G}{r^{2}(m_{s},t)}\int^{m_{s}}_{0}dm_{s}'\mu(m_{s}',t)~.
\label{eq:dim2}
\ee

\subsection{Numerical calculation}

To solve our system of coupled partial differential equations numerically, we divided the collapsing sphere into $N$ shells, initially with equal masses and with the $n$th such shell having a time-dependent mass $\Delta m_{n}(t)$ and radius $r_{n}(t)$.  If $n>1$, then the spatially discretized equations of motion for this shell are
\ba
\frac{d^{2}r_{n}(t)}{dt^{2}}&=&-\frac{G}{r_{n}(t)^{2}}\sum_{i=1}^{n}\Delta m_{i}(t)\nonumber\\
\frac{d\Delta m_{n}(t)}{dt}&=&-\frac{3\sigv}{4\pi m}\Delta m_{n}(t)\left(\frac{dm_{s}}{dr^{3}}\right)_{n}\nonumber\\
&\approx& -\frac{3\sigv}{4\pi m}\left(\frac{\Delta m_{n}(t)^{2}}{r_{n}(t)^{3}-r_{n-1}(t)^{3}}\right)~,
\ea
where the central ``shell" ($n=1$) is treated as a homogenous sphere as in the previous section, and thus will have the following equations of motion:
\ba
\frac{d^{2}r_{1}(t)}{dt^{2}}&=&-\frac{G}{r_{1}(t)^{2}}\Delta m_{1}(t)\nonumber\\
\frac{d\Delta m_{1}(t)}{dt}&=&-\frac{3\sigv}{4\pi m}\left(\frac{\Delta m_{1}(t)^{2}}{r_{1}(t)^{3}}\right)~.
\ea
Our requirement in the previous subsection that $\alpha$ be large also requires that this innermost shell is significantly larger than its annihilation radius at $t=0$.  For $\alpha=3/4\pi\times 10^{6}$, $r_{1}(t=0)$ is one hundred times greater than the annihilation radius for a homogenous sphere of mass $\Delta m_{1}(t=0)$.

In order to nondimensionalize this system, we choose convenient units as before with the homogenous case; let the mass unit be the initial shell mass $\Delta m_{0}=M_{0}/N$ and let the length unit be the previously-defined annihilation length for the central shell
\be
R_{ann}=\left(\frac{3\sigv}{4\pi mG}\right)^{2/3}\left(\frac{GM_{0}}{N}\right)^{1/3}~.
\ee
We also choose a corresponding time unit
\be
t_{c}=\sqrt{\frac{R_{ann}^{3}}{G\Delta m_{0}}}
\ee
such that the equations take on a simpler form in terms of the new dimensionless variables $\Delta \hat{m}_{n}=\Delta m_{n}/\Delta m_{0}$, $\hat{r}_{n}=r_{n}/R_{ann}$ and $\tau=t/t_{c}$:
\ba
\frac{d^{2}\hat{r}_{n}(\tau)}{d\tau^{2}}&=&-\frac{1}{\hat{r}_{n}(\tau)^{2}}\sum_{i=1}^{n}\Delta\hat{m}_{i}(\tau)\nonumber\\
\frac{d\Delta\hat{m}_{n}(\tau)}{d\tau}&\approx&-\frac{\Delta \hat{m}_{n}(\tau)^{2}}{\hat{r}_{n}(\tau)^{3}-\hat{r}_{n-1}(\tau)^{3}}
\ea
for $n>1$ and
\ba
\frac{d^{2}\hat{r}_{1}(\tau)}{d\tau^{2}}&=&-\frac{\Delta\hat{m}_{1}(\tau)}{\hat{r}_{1}(\tau)^{2}}\nonumber\\
\frac{d\Delta\hat{m}_{1}(\tau)}{d\tau}&\approx&-\frac{\Delta \hat{m}_{1}(\tau)^{2}}{\hat{r}_{1}(\tau)^{3}}
\ea
for $n=1$. These equations were integrated in time using a fourth order Runge-Kutta routine similar to that in the above section, with the initial time $\tau=0$, initial shell masses $\Delta \hat{m}_{n}=1$ and initial radii given by equation (\ref{rnoann}) to be
\be
\hat{r}_{n}(\tau=0)=\left(\frac{4\pi\alpha n}{3}\right)^{1/3}\left(\gamma\left(\frac{n}{N}\right)^{b}+1\right)^{2/3}~.
\ee
The initial infall velocities were chosen to be
\be
\left(\frac{d\hat{r}_{n}}{d\tau}\right)_{t=0}=-\sqrt{\frac{2}{\hat{r}_{n}}\sum^{i=1}_{n}\Delta \hat{m}_{i}}~,
\ee
which corresponds to zero energy infall.

\begin{figure}[h]
\includegraphics[width=\picwidth,height =9cm ]{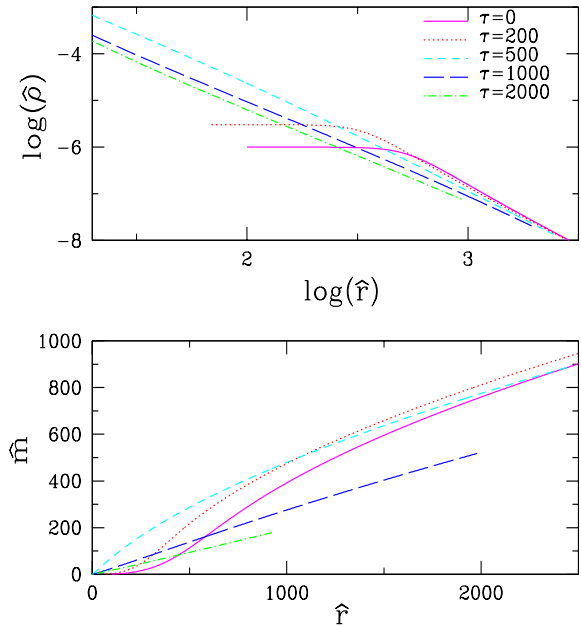}
\caption{Data for a simulation run in which a black hole does not form, with $b=2$, $N=1000$, $\alpha=3/4\pi\times 10^{6}$ and $\beta=100$.  The top plot shows the log of the dimensionless density $\hat{\rho}$ versus the log of the radius $\hat{r}$ for several times $\tau=0, 200, 500, 1000$ and $2000$.  The bottom plot shows the dimensionless mass enclosed versus radius for the same $\tau$ values.} 
\label{onedim1} 
\end{figure}

As we would expect from the homogeneous case, each shell reached both zero size and zero mass at the same time in the absence of black hole formation, i.e. for clouds which annihilate completely.  Some results for this simulation are shown in Figure~\ref{onedim1} for $b=2$, $\alpha=3/4\pi\times10^6$, $\beta_{i}=100$ and $N=1000$.  We chose to use one thousand shells for the simulation because this $N$ provided sufficient spatial resolution while still allowing the code to run within a reasonable amount of time.  

It is important to note that, for this set of parameters, there appears to be a transition around a time $\tau\sim500$.  This is simply the annihilation time $t_{ann}$, defined in equation (\ref{anntime}) to be the time at which the center reaches the annihilation density:
\be
t_{ann}=t_{0}-\frac{1}{\sqrt{6\pi G\rho_{ann}}}=t_{0}\left(1-\sqrt{\frac{3}{8\pi\alpha}}\right)~;
\ee
using $t_{0}=\sqrt{\alpha/6\pi}(\sigv/Gm)$, $\alpha=3/4\pi\times 10^{6}$ and our nondimensional time variable, we find that $\tau_{ann}\equiv t_{ann}/t_{c}\approx 471$.  This is the same value as we see in the numerical solution, and therefore we can conclude that neglecting annihilation effects until reaching the critical length scale is a good approximation for this choice of parameters.

\subsection{Black hole formation}

Our main goal is to generalize our investigation of black hole formation in uniform density collapse to non-uniform collapse.  To that end, at each time step, we check to see if any shell satisfies the condition
\be
\frac{2Gm_{s}}{r}\geq 1~;
\ee
in terms of our nondimensional variables and spatial discretization, this condition becomes
\be
\left[\frac{1}{2N^{2}}\left(\frac{M_{0}}{M_{ann}}\right)^{2}\right]^{1/3}\left(\frac{1}{\hat{r}_{n}}\sum^{i=1}_{n}\Delta \hat{m}_{i}\right)\geq 1~.
\ee
If this condition is true for a given shell, then we stop integrating the equations of motion for all shells inside it.  In other words, we find the location of the event horizon and shut off the annihilation and collapse of the mass inside it.  Then, if a black hole does form, there will be an initial seed hole that subsequently will accrete mass from the remaining (annihilating) matter outside.  If a black hole does not form, then all of the shells will eventually annihilate away until no mass is left.  Again, we shall see that the initial cloud mass $M_{0}$ must be above some critical minimum value, $M_{c}$, in order for the cloud to collapse to a black hole.

\begin{figure}[h]
\includegraphics[width=\picwidth,height =9cm ]{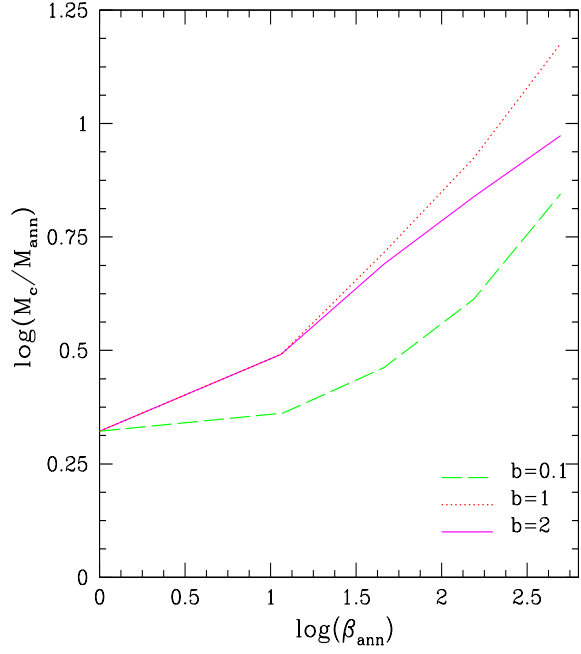}
\caption{Simulation data plot of $\log(M_{c}/M_{ann})$ versus $\log(\beta_{ann})$ for $b=0.1, 1$ and $2$.} 
\label{betas} 
\end{figure}

In general, the mass scale $M_{c}$ depends on $\beta_{ann}$ and $b$.  To determine $M_{c}$ for a given choice of $b$ and $\beta_{ann}$, we increased our inputted value of the mass ratio $M_{0}/M_{ann}$ until a black hole could form.  (Of course, $\beta_{ann}=1$ reproduced the results found for the homogenous case, independent of the value of $b$.)  As we increased the contrast $\beta_{ann}$, we found that this minimum mass, or $M_{c}$, also increased via a rough power law.  These results, for several $b$ values, are plotted in Figure~\ref{betas}.  Note that the exponent of the power law depends on $b$.

We can estimate the critical masses in Figure~\ref{betas} by using the LL solution and simultaneously requiring that a black hole forms before the annihilation time $t_{ann}$.  In the LL solution,
\be
\phi\equiv\frac{2Gm_{s}}{r}=\left[\frac{4Gm_{s}}{3\left(Bm_{s}^{b}+t_{0}-t\right)}\right]^{2/3}~,
\ee
and this has a maximum value when
\be
m_{s}=\left(\frac{t_{0}-t}{B\left(b-1\right)}\right)^{1/b}~;
\label{phimax}
\ee
note that this becomes singular as $b\rightarrow 1$, and therefore we will assume $b>1$ for the time being.  Inserting equation (\ref{phimax}) into our equation for $\phi$ gives us the maximum value
\be
\phi_{max}\equiv\left[\frac{4G\left(b-1\right)^{1-1/b}}{3bB^{1/b}\left(t_{0}-t\right)^{1-1/b}}\right]^{2/3}~;
\ee
$\phi_{max}=1$ at the time of initial black hole formation, $t_{BH}$, i.e. when
\be
t=t_{BH}\equiv t_{0}-\left(\frac{4G}{3}\right)^{\frac{b}{b-1}}\left(\frac{b-1}{B^{\frac{1}{b-1}}b^{\frac{b}{b-1}}}\right)~.
\ee
In order to produce a black hole before annihilation becomes important, we want $\phi_{max}=1$ and $t<t_{ann}$; this happens if
\be
B<\left(\frac{4G}{3}\right)^{b}\frac{\left(b-1\right)^{b-1}}{b^{b}\left(t_{0}-t_{ann}\right)^{b-1}}~.
\ee
Using equation (\ref{B}), we can rewrite this condition as
\be
M_{0}>\frac{3b\left(t_{0}-t_{ann}\right)}{4G\left(b-1\right)^{1-1/b}}\left[\sqrt{\left(\frac{b+1}{2b+1}\right)^{2}+\frac{\beta_{ann}-1}{2b+1}}-\left(\frac{b+1}{2b+1}\right)\right]^{1/b}\equiv M_{c}~.
\ee
For large $\beta_{ann}$, this yields the power law $M_{c}\propto \beta_{ann}^{1/2b}$.

If $b\leq 1$, the situation is somewhat different.  Note that, for any b value,
\be
\frac{\partial}{\partial m_{s}}\left(\frac{2Gm_{s}}{r}\right)\propto t_{0}-t+\left(1-b\right)Bm_{s}^{b}~,
\ee
which is positive for $b\leq 1$ and $t_{0}-t>0$, indicating that $\phi$ is an increasing function of $m_{s}$ under these circumstances.  In that case, $\phi$ is largest at the edge, where $m_{s}=M_{0}$.  Following the same line of reasoning as for the $b>1$ case, requiring that $\phi_{max}=1$ and $t<t_{ann}$, we find that
\be
M_{c}=\frac{3\left(t_{0}-t_{ann}\right)}{4G}\left(\sqrt{\left(\frac{b+1}{2b+1}\right)^{2}+\frac{\beta_{ann}-1}{2b+1}}+\frac{b}{2b+1}\right)
\label{mc}
\ee
for the $b\leq 1$ case.  Using the data in Figure~\ref{betas}, we have found that this estimate is too large by a factor $\sim 2$ for the $b=0.1$ case, even though the data do follow the predicted power law $M_{c}\propto \beta_{ann}^{1/2}$ for larger values of $\beta_{ann}$; in general, the estimate of equation (\ref{mc}) is too large by factors of order unity for $b$ values less than one.  Hence, for $b\leq 1$, we can conclude that it is possible for black holes to form at a later time $t=t_{ann}+\Delta t$, such that $M_{c}$ is reduced by a factor of $(t_{0}-t_{ann}-\Delta t)/(t_{0}-t_{ann})$.

The simulation also tracked the masses of both the initial seed black holes and the end-state black holes that they eventually evolved to.  Theoretically, for $b\leq 1$, we expect that the initial black hole will have a mass of order $M_{0}$, and this prediction was verified by the simulation.  For $b>1$, we should find the initial black hole mass from equation (\ref{phimax}):
\be
M_{BH}=\left(\frac{t_{0}-t_{BH}}{B\left(b-1\right)}\right)^{1/b}~;
\ee
plugging in what we know for $t_{BH}$ gives us
\be
M_{BH}=\left(\frac{M_{0}^{b}}{bM_{ann}}\left[\sqrt{\left(\frac{b+1}{2b+1}\right)^{2}+\frac{\beta_{ann}-1}{2b+1}}-\frac{b+1}{2b+1}\right]^{-1}\right)^{\frac{1}{b-1}}\propto M_{ann}\left(\frac{M_{0}}{M_{ann}}\right)^{\frac{b}{b-1}}
\ee
which agrees with the simulation data.

Finding the final black hole mass is equivalent to finding the total mass change due to annihilation, which we will call $\Delta M$.  Suppose that a black hole forms with a mass that is $M_{BH}$ initially.  The mass density at this time is $\rho_{BH}\sim3/32\pi G^{3}M_{BH}^{2}$, so annihilation proceeds at a rate $\approx\rho_{BH}\sigv M_{BH}/m\sim 3\sigv/32\pi G^{3}mM_{BH}\sim M_{ann}/GM_{BH}$.  Up to numerical factors, we therefore expect a peak annihilation luminosity $-\dot{M}\sim M_{ann}/GM_{BH}$ and a total mass loss to annihilation $\Delta M\sim -M_{ann}$.  Thus, whenever a black hole forms, a similar amount of energy is released in the form of annihilation products.

To study this more completely, we consider collapsing clouds well above the critical mass for black hole formation. The easiest case is the collapse of a homogenous cloud ($b=0$).  If we assume that the initial mass is significantly above $M_{c}$, then we can assume that annihilation will not affect the trajectories of the shells as they fall inward.  For the uniform case
\be
\rho(t)=\frac{1}{6\pi G\left(t_{0}-t\right)^{2}}
\ee
and a black hole forms when $t_{0}-t_{BH}=4GM_{0}/3$.  The total mass annihilated is
\ba
\Delta M&=&-\frac{\sigv M_{0}}{6\pi Gm}\int^{t_{BH}}_{-\infty}\frac{dt}{\left(t_{0}-t\right)^{2}}=-\frac{\sigv M_{0}}{6\pi Gm}\left(\frac{1}{t_{0}-t_{BH}}\right)\nonumber\\
&=&-\frac{\sigv}{8\pi G^2m}=-\frac{2}{3}M_{ann}~.
\ea
Therefore, the total mass that is lost to annihilation in the homogenous case is of order $M_{ann}$, for any initial mass that is greater than the minimum mass $M_{c}$.  

All cases with $b\leq 1$ are similar to the $b=0$ case in that the event horizon first forms at the outer edge.  In this case, annihilation ends when
\be
t_{0}-t_{BH}=\frac{4M_{0}}{3}-BM_{0}^{b}~;
\ee
using the LL solution for $\rho(m_{s},t)$ we find
\be
\Delta M=-\frac{\sigv}{12\pi bGmBM_{0}^{b-1}}\int_{0}^{1}\frac{dx}{x^b}\ln\left[\frac{\left(2b+1\right)BM_{0}^{b}x^{b}-BM_{0}^{b}+4M_{0}/3}{BM_{0}^{b}x^{b}-BM_{0}^{b}+4M_{0}/3}\right]~.
\label{smallb}
\ee
In the limit that $b$ goes to zero, equation (\ref{smallb}) yields the same answer as before: $\Delta M\rightarrow-2M_{ann}/3$.

For $b>1$, the situation is more complicated because annihilation continues after the initial black hole formation.  For this case, it is useful to think in terms of the luminosity as a function of time, which is
\ba
L(t)\equiv-\frac{d\Delta M}{dt}&=&\frac{\sigv}{m}\int^{M_{0}}_{M_{min}(t)}\rho(m_{s},t)dm_{s}\nonumber\\
&=&\frac{\sigv}{6\pi Gm}\int^{M_{0}}_{M_{min}(t)}\frac{dm_{s}}{\left(Bm_{s}^{b}+t_{0}-t\right)\left[\left(2b+1\right)Bm_{s}^{b}+t_{0}-t\right]}~,
\label{deltam2}
\ea
where $M_{min}(t)$ is the mass within the event horizon, which doesn't contribute to the luminosity.  Let $t_{BH}$ be the time when a black hole first forms; then for $t<t_{BH}$, we have $M_{min}(t)=0$, and for $t>t_{BH}$ we have
\be
\frac{4GM_{min}(t)}{3}=B\left[M_{min}(t)\right]^{b}+t_{0}-t~.
\label{mmin}
\ee
For very early times, when $t<t_{BH}$ and $t_{0}-t>>BM_{0}^{b}$, there is an initial phase where we can approximate
\ba
\left(\frac{d\Delta M}{dt}\right)_{early}&=&-\frac{\sigv}{6\pi Gm}\int^{M_{0}}_{0}\frac{dm_{s}}{\left(Bm_{s}^{b}+t_{0}-t\right)\left[\left(2b+1\right)Bm_{s}^{b}+t_{0}-t\right]}\nonumber\\
&\approx&-\frac{\sigv}{6\pi m}\left[\frac{M_{0}}{\left(t_{0}-t\right)^{2}}\right]~.
\ea
For smaller $t_{0}-t$, but still before black hole formation, we have a another phase where
\be
\left(\frac{d\Delta M}{dt}\right)_{t<t_{BH}}=-\frac{\sigv}{6\pi Gm\left(t_{0}-t\right)^{2}}\int^{M_{0}}_{0}\frac{dm_{s}}{\left[Bm_{s}^{b}/\left(t_{0}-t\right)+1\right]\left[\left(2b+1\right)Bm_{s}^{b}/\left(t_{0}-t\right)+1\right]}~;
\ee
if we define $x^{b}\equiv Bm_{s}^{b}/(t_{0}-t)$ and assume that $(M_{0}/M_{BH})^{b}>>1$, then this becomes
\be
\left(\frac{d\Delta M}{dt}\right)_{t<t_{BH}}\approx -\frac{b^{2}M_{ann}}{2\left(b-1\right)^{2-1/b}GM_{BH}}\left(\frac{t_{0}-t_{BH}}{t_{0}-t}\right)^{2-1/b}\int^{\infty}_{0}\frac{dx}{\left(x^{b}+1\right)\left[\left(2b+1\right)x^{b}+1\right]}~.
\ee
The $d\Delta M/dt$ of this second phase increases with time until it ends right before $t=t_{BH}$, and so we get a peak luminosity of
\be
\dot{M}_{peak}\approx\frac{b^{2}M_{ann}}{2\left(b-1\right)^{2-1/b}GM_{BH}}\int^{\infty}_{0}\frac{dx}{\left(x^{b}+1\right)\left[\left(2b+1\right)x^{b}+1\right]}~.
\ee
Right at $t=t_{BH}$, we know that $M_{min}=M_{BH}$, and so then we have
\be
\left(\frac{d\Delta M}{dt}\right)_{t=t_{BH}}\approx -\frac{b^{2}M_{ann}}{2\left(b-1\right)^{2-1/b}GM_{BH}}\int^{\infty}_{\left(b-1\right)^{-1/b}}\frac{dx}{\left(x^{b}+1\right)\left[\left(2b+1\right)x^{b}+1\right]}~,
\ee
and thus, in our simple model, $d\Delta M/dt$ drops by a factor of order one right after a black hole forms.  

The total mass that is annihilated before $t_{BH}$, which we call $\Delta M_{1}$, is simply
\be
\Delta M_{1}=\int_{-\infty}^{t_{BH}}dt\frac{d\Delta M(t)}{dt}~;
\label{delta1}
\ee
the mass annihilated from the phase where $d\Delta M/dt\approx (d\Delta M/dt)_{early}$ is approximately
\be
\Delta M_{early}\approx -\frac{2bM_{ann}}{3}\left(\frac{M_{BH}}{M_{0}}\right)^{b-1}
\ee
which is very small for $M_{0}/M_{BH}>>1$, and therefore we will neglect this term; then equation (\ref{delta1}) becomes
\be
\Delta M_{1}\approx\int_{-\infty}^{t_{BH}}dt\left(\frac{d\Delta M(t)}{dt}\right)_{t<t_{BH}}\approx -\frac{2b^{2}M_{ann}}{3\left(b-1\right)^{2-1/b}}\int^{\infty}_{0}\frac{dx}{\left(x^{b}+1\right)\left[\left(2b+1\right)x^{b}+1\right]}~,
\ee
where we took the upper mass limit to infinity, which is justified for $M_{0}>>M_{BH}$ and $b>1$.

After $t_{BH}$, we need to use the more general equations (\ref{deltam2}) and (\ref{mmin}), but for very late times, when $t-t_{0}>>GM_{min}(t)$, we can approximate
\be
M_{min}(t)\approx\left(\frac{t-t_{0}}{B}\right)^{1/b}\left(1+\epsilon\right)~,
\ee
where
\be
\epsilon=\frac{4M_{min}(t)}{3b\left(t-t_{0}\right)}~;
\ee
then, in this very late phase, we can define the integration variable $x\equiv m_{s}/M_{min}(t)$ to get
\ba
\left(\frac{d\Delta M}{dt}\right)_{late}&\approx&-\frac{\sigv M_{min}(t)}{6\pi Gm\left(t-t_{0}\right)^{2}}\int^{\infty}_{1}\frac{dx}{\left[\left(1+\epsilon\right)x^{b}-1\right]\left[\left(2b+1\right)x^{b}-1\right]}\nonumber\\
&\approx&-\frac{b^{2}M_{ann}}{2\left(b-1\right)^{1-1/b}GM_{BH}}\left(\frac{t_{0}-t_{BH}}{t-t_{0}}\right)^{2-1/b}\ln\left(\frac{t-t_{0}}{t_{0}-t_{BH}}\right)~.
\ea

The total mass annihilated after $t_{BH}$, which we call $\Delta M_{2}$, is given by
\be
\Delta M_{2}=-\frac{\sigv}{6\pi Gm}\int_{M_{BH}}^{M_{0}}dm_{s}\int^{t_{f}(m_{s})}_{t_{BH}}\frac{dt}{\left(Bm_{s}^{b}+t_{0}-t\right)\left[\left(2b+1\right)Bm_{s}^{b}+t_{0}-t\right]}~,
\ee
where $t_{f}(m_{s})$ is the time at which a mass shell with a coordinate $m_{s}$ reaches the event horizon:
\be
\frac{4Gm_{s}}{3}=Bm_{s}^{b}+t_{0}-t_{f}(m_{s})~;
\ee
if we use the integration variable $x\equiv m_{s}/M_{BH}$, we find that
\be
\frac{\Delta M_{2}}{M_{ann}}\approx -\frac{1}{3}\int^{\infty}_{1}\frac{dx}{x^{b}}\ln\left\{\left(2x^{b-1}+1\right)\left[\frac{\left(2b+1\right)x^{b}+b-1}{x^{B}+b-1}\right]\right\}
\ee
for $M_{0}>>M_{BH}$ and $b>1$.

\begin{figure}[h]
\includegraphics[width=\picwidth,height =9cm ]{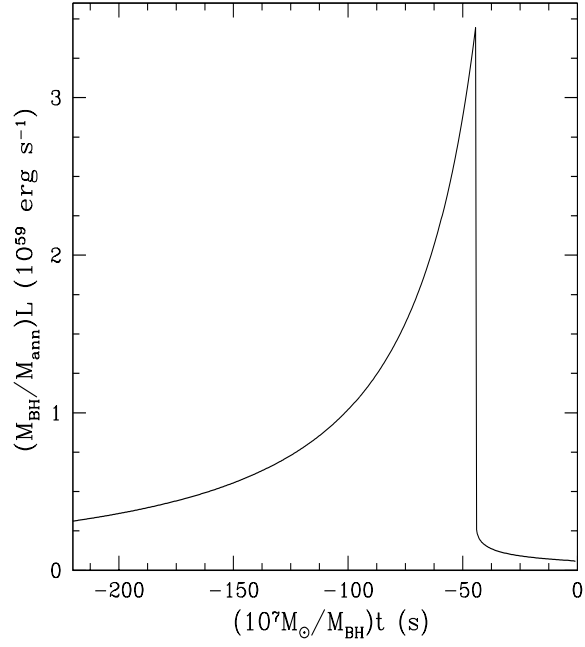}
\caption{Plot of $L(t)$ versus $t$ for $b=2$.} 
\label{lightcurve} 
\end{figure}

We can integrate the above equations numerically to get $\Delta M_{1}$, $\Delta M_{2}$ and $\dot{M}_{peak}$, and we show the results for several values of $b$ in Table~\ref{table1}.
\begin{table}[h]
\caption{Data for several values of $b$}
\begin{ruledtabular}
\begin{tabular}{cccc}
$b$ & $\Delta M_{1}/M_{ann}$ & $\Delta M_{2}/M_{ann}$ & $G\dot{M}_{peak}M_{BH}/M_{ann}$ \\
\hline
2 & -1.29 & -1.08 & -0.97 \\
3 & -1.01 & -0.57 & -0.76 \\
4 & -0.91 & -0.39 & -0.68 \\
\end{tabular}
\end{ruledtabular}
\label{table1}
\end{table}
We also provide a sample lightcurve in Figure~\ref{lightcurve}, which shows the scaled luminosity $LM_{BH}/M_{ann}$ as a function of the scaled time $t(\rm s)(10^{7}M_{\odot}/M_{BH})\approx (50~{\rm s})(t/GM_{BH})$.  Our calculated light curve ignores the effects of opacity and general relativity.  The former should be relatively unimportant when a black hole forms, as was discussed for the uniform case in Section IIB.  (When $b<1$, the uniform case is a good approximation, and when $b>1$, most of the optical depth for annihilation products is near the hole.)  General relativity would tend to lower the luminosity peak and smooth it out.  Within factors of $\sim 2$ or $3$, we still expect a lightcurve resembling Figure~\ref{lightcurve}, and a total annihilated mass $\sim M_{ann}$.

\section{Discussion}

\subsection{P-Wave Annihilation}

P-wave annihilation has $\sigv=\sigma_0\vrms$. In the early
Universe, $\vrms=T/m$, and following \cite{KT} we find
\be
\sigma_0\approx 1.7\times 10^{-7}~\gev^{-2}~\left({x_f^2
\over 10^3\omcdm\gstars/\sqrt{\gstar}}\right)~,
\ee
where $x_f\approx C-{3\over 2}\ln(\ln C)$ with
$C\approx 0.076(g/\sqrt{\gstar})m\sigma_0/\sqrt{G}$.
Assuming that thermal particles scatter from the CDM
particles with a cross-section $f_t\sigma_0(T/m)^2$, 
we find that the CDM remains in kinetic equilibrium
until
\be
T_K\approx 1.6\times 10^{-3}~\gev~{\gstars^{1/4}
[m(\gev)]^{3/4}\over g_t^{1/4}f_t^{1/4}}
\left({10^3\omcdm\over x_f^2}\right)^{1/4}~;
\ee
consequently
\be
K_{CDM}\approx 10^9~\gev^{-8/3}{g_t^{1/4}f_t^{1/4}\over
[m(\gev)]^{7/4}\gstars^{1/4}\gstarsk^{2/3}}
\left({x_f^2\over 10^3\omcdm}\right)^{1/4}\left({0.15
\over\omcdm}\right)^{2/3}~.
\ee
As long as the collapse proceeds adiabatically, 
$\vrms=K_{CDM}\rho^{2/3}$, and
for pure p-wave annihilation we find
\be
\Mann\approx{2.4\times 10^8\msun\over[m(\gev)]^{33/28}}
~\left({g_t^{3/28}f_t^{3/28}\gstar^{3/14}\over
\gstars^{15/28}\gstarsk^{2/7}}\right)
\left({x_f^2\over 10^3\omcdm}\right)^{15/28}\left({0.15
\over\omcdm}\right)^{2/7}~.
\ee
For example, if $m=50~\gev$ and $g_t\approx\gstar\approx
\gstars\approx\gstarsk\approx 50$, we find
$\Mann\approx 10^6~f_t^{3/28}\msun$. Not suprisingly, the
value of $\Mann$ is considerably smaller than for s-wave
annihilation, under the assumption of adiabatic collapse:
p-wave annihilation is less efficient, and therefore the
cloud must reach higher density to annihilate, which
lowers the critical mass for black hole formation. 

In actuality, we do not expect the cloud to
collapse adiabatically, and the critical mass is
closer to the value for s-wave annihilation at
$\sigv=\sigma_0$. The reason is that $\rhoann$
is large enough that clouds become opaque
to their annihilation products {\it before} they
either annihilate or collapse catastrophically. Defining
$\rhonead$ to be the density at which the optical depth is
one in adiabatic collapse, we find
\be
\rhonead=\rhoann f_s^{-3/2}(K\rhoann^{2/3})^{3/2}
\left({\Mann\over M}\right)^{1/2}=\rhobh f_s^{-3/2}
(K\rhoann^{2/3})^{3/2}
\left({M \over\Mann}\right)^{3/2}~;
\ee
since
\be
K_{CDM}\rhoann^{2/3}
\approx{1.2\times 10^{-3}\over[m(\gev)]^{5/28}}
\left({g_t^{3/28}f_t^{3/28}\gstars^{13/28}\over
\gstarsk^{2/7}}\right)\left({10^3\omcdm\over x_f^2}
\right)^{13/28}\left({0.15\over\omcdm}\right)^{2/7}~,
\ee
we see that the clouds become opaque before reaching their
final destinies.

Once the clouds become opaque, they start to heat up. To
understand what happens qualitatively, we introduce new
density and mass scales
\ba
\rho_1&=&{8\pi Gm^2\over 3f_s^2\sigma_0^2}=\rhoann\left(
{K\rhoann^{2/3}\over f_s}\right)^2\nonumber\\
M_1&=&\sqrt{3\over 32\pi G^3\rho_1}={3f_s\sigma_0\over
16\pi G^2m}=\Mann\left({f_s\over K\rhoann^{2/3}}\right)
~,
\ea
in terms of which we find
\ba
\tau&=&\left({\rho\over\rho_1}\right)^{2/3}\left({M\over
M_1}\right)^{1/3}\nonumber\\
\vff^2&=&{2GM\over R}=\left({\rho\over\rho_1}\right)^{1/3}
\left({M\over M_1}\right)^{2/3}~;
\ea
we can eliminate the density from these equations to find
\be
\vff^2=\tau^{1/2}\left({M\over M_1}\right)^{1/2}~.
\ee
Thus, we see that clouds with $M\geq M_1$ collapse to 
black holes before becoming opaque, and vice-versa.
We also see that during adiabatic collapse,
\ba
{\Dm\over M}&\equiv&{\Gamma_{\rm ann}\over H}
=\left({K\rhoann^{2/3}\over f_s}\right)^{7/3}\tau^{7/4}
\left({M_1\over M}\right)^{7/12}\nonumber\\
K\rho^{2/3}&=&f_s\left({K\rhoann^{2/3}\over f_s}\right)^{7/3}
\tau\left({M_1\over M}\right)^{1/3}={f_s\over\tau^{3/4}}
\left({M\over M_1}\right)^{1/4}{\Dm\over M}~.
\label{adeqns}
\ea
Note that $\Dm/M$ is the fractional mass annihilated in one free-fall
time up to a factor of order unity.

Consider the fate of a cloud with $M/M_1\lesssim 1$, but
not $\ll 1$ (we shall estimate how much smaller shortly).
Once the cloud reaches $\tau>1$, the relativistic products
of the annihilation do not escape freely, and deposit a 
substantial fraction of their total energy into kinetic
energy of the massive particles. Because the optical 
depth for the relativistic particles is proportional
to $(E/m)^2$, after a relativistic particle scatters once,
thereby losing about half of its original energy on average,
its optical depth to escape drops by a substantial factor;
hence, as long as $\tau$ is not too large, relativistic
particles are not trapped. The massive particles that
scatter the relativistic particles are heated to semi-relativistic
kinetic energies i.e. $\sim m/2$. For these, however, the
scattering cross section remains $\approx\sigma_0$; moreover,
on each scatter off of one of the colder massive particles
that dominate the cloud, these hot massive particles lose
of order half of their kinetic energy. As a result, they
continue to slow down, and it is likely that they become
trapped inside the cloud. The second of equations (\ref{adeqns})
shows that $\Dm/M\gtrsim K\rho^{2/3}$ for clouds masses
$\lesssim M_1$ at $\tau\gtrsim 1$, so the cloud heats up
to $\vrms=\eta\Dm/M$, with $\eta\lesssim 1/2$.
Thus, we expect the internal energy of the cloud to rise
rapidly to $\eta\Dm/M$ as it continues to collapse to 
larger optical depth.

Since the internal energy of the cloud rises as it
annihilates, the annihilation equation is altered to
\be
{dM\over dt}=-{d\Dm\over dt}\approx-{\eta\rho\sigma_0
\Dm\over m}~;
\ee
for $\Dm\lesssim M$ we find
\be
\ln\Dm=\ln\Dm_i+{2\eta\over 3}\left[\left({\rho\over\rho_1}
\right)^{1/2}-\left({\rho_i\over\rho_1}\right)^{1/2}\right]~.
\ee
Taking $\rho_i$ and $M_{i}$ to correspond to $\tau=1$, we find that
$\Dm\to M$ when the density must increase to
\be
\left({\rho\over\rho_1}\right)^{1/2}\approx{7\over 2\eta}\ln\left({f_s\over
K\rhoann^{2/3}}\right)-{7\over 8\eta}\ln\left({M_1\over M}\right)
+\left({M_1\over M}\right)^{1/4}~,
\ee
for the cloud to annihilate away. (The last stages of the cloud's
collapse could deviate somewhat from free-fall, as the pressure
builds up to values comparable to $\rho\vff^2$, but this phase most
likely is only short-lived.) Thus, we estimate that the critical
mass for a black hole to form is of order 
\be
M_{c, p}={2\eta M_1\over 7\ln(f_s/K\rhoann^{2/3})}
={2\eta f_s\Mann\over 7K\rhoann^{2/3}\ln(f_s/K\rhoann^{2/3})}~;
\ee
for example, if $m=50~\gev$ and $g_t\approx\gstar\approx
\gstars\approx\gstarsk\approx 50$, we find that
$M_{c, p}\approx 10(2\eta)\Mann$ or of order
$10^7(2\eta)\msun$. We note that since $M_1$ is of order
$\Mann$ for s-wave annihilation with $\sigv=\sigma_0$,
the critical mass for black
hole formation for p-wave annihilation is only smaller by 
a factor of $\sim 7\ln(f_s/K\rhoann^{2/3})/2\eta$, not by the factor
of $\sim K\rhoann^{2/3}$ deduced under the false assumption of
adiabatic collapse.

It is also possible that the CDM particles can annihilate
via both s-wave and p-wave channels with p-wave dominant
in the early Universe but s-wave dominant during the early
stages of the cloud collapse.
If the s-wave annihilation has $\sigv=\eps\sigma_0$, then
it dominates until $\Dm/M\approx\eps$, whereupon the
p-wave channel takes over. In this situation, we expect
to get a critical mass for black hole formation that is
smaller than for s-wave annihilation with $\sigv=\sigma_0$
by a logarithmic factor of order $(3/2\eta)\ln\eps^{-1}$.

\subsection{Conclusions}

Our calculations show that the collapse of a cloud of annihilating
matter to a black hole is only possible for masses above
a critical mass, $M_c$, that is of order $10^7-10^8\msun$
for plausible cold dark matter candidates. For a radial density profile, the critical mass depends on the parameters of the profile; for example, we found that $M_{c}/M_{ann}\sim\beta_{ann}^{1/2b}$ for an LL profile with the exponent $b$ \cite{LL}, where $\beta_{ann}$ is the density contrast between the center and the edge when annihilation becomes important. The conditions
for collapse studied here are highly idealized since
we assumed that the clouds are exactly spherical and cold,
at least initially. Internal shearing motions induced
by tidal perturbations in the early Universe would violate
these assumptions. However, it is conceivable that the
collapse of larger, sheared structures could very rarely
lead to the formation of bound
fragments similar to what we have assumed here \cite{LoebRasio,VMSV}. In any event, it is likely
that we have assumed conditions most favorable for forming
a black hole from CDM collapse, so our lower bounds on the
masses of black holes that can form in this way are probably,
if anything, underestimates of reality. It is remarkable,
although perhaps completely fortuitous,
that the masses estimated here -- which are determined by
fundamental constants of Nature -- are even close to the
sorts of black hole masses observed astronomically \cite{BH}. 

When a cloud collapses to a black hole, annihilation releases energy $\sim M_{ann}$ in a time $\sim GM_{BH}$, where $M_{BH}>M_{ann}$ is the mass of the black hole when collapse first occurs.  The peak luminosity resulting from the annihilation is therefore $\sim M_{ann}/GM_{BH}$, and much of the annihilation occurs in a prodigious, relatively short-lived burst.  Thus the formation of a black hole in the collapse of annihilating CDM would be a spectacular event: for s-wave annihilation,
approximately $M_{ann}\sim 2\times 10^{64}
[m(\gev)]^{-1}(x_f\sqrt{\gstar}/15\gstars)$ ergs would
be released in a time of order $10^5[m(\gev)]^{-1}
(x_f\sqrt{\gstar}/15\gstars)(M_{BH}/M_{ann})$ seconds, with a peak luminosity of $4\times 10^{59}(M_{ann}/M_{BH})$ ergs per second; for $m=50$ GeV,
the energy released is of order $7\times 10^{61}$ ergs
over about $380(M_{BH}/M_{ann})$ seconds.  For comparison, note that $\frac{1}{2}(10^{11}M_{\odot})(300~{\rm km~s^{-1}})^{2}\approx G(10^{11}M_{\odot})^{2}/10~{\rm kpc}\approx 10^{59}$ ergs; i.e. the energy released in forming a single annihilating CDM black hole is comparable to, if not larger than, the binding energy of an entire galaxy.
 
What an observer could detect from
such events depends on what annihilation products form,
but even if the efficiency of generating electromagnetic
radiation is $\lesssim 10^{-8}$ of the total energy
released, they would be as
bright as extragalactic gamma ray bursts.
Any direct production of standard
model particles in the annihilation should produce a pair
fireball with little or no net baryon content,
even for extremely low branching ratios for
$mm\to({\rm standard~model~pairs})$. For example,
if there is no direct production of $e^\pm$ pairs or quarks, but a fraction $f_\nu$ of the annihilation energy goes into
neutrinos and antineutrinos of energy $m$, then we estimate an efficiency of
producing $e^\pm$ pairs of about $(0.01-0.1)f_\nu
G_F^2m^2/\sigv\approx (0.001-0.01)
f_\nu[m(\gev)]^2(x_f\sqrt{\gstar}/15\gstars)^{-1}$
for s-wave annihilation,
and perhaps a bit smaller for p-wave annihilation; this would
lead to a total energy of order $(0.02-0.002)\times 10^{64}
f_\nu^2m(\gev)$ ergs \cite{GDN}. Thus,
a tiny branching ratio for $mm\to\nu{\overline{\nu}}$
would result in the formation of an extremely energetic pair
fireball, even without any direct pair production via 
$mm\to e^+e^-$ or $mm\to q{\overline{q}}$: if $f_\nu\sim
10^{-5}$, we estimate a total pair fireball energy of order
$10^{51}-10^{52}m(\gev)$ ergs. Acceptable branching ratios for direct production
of $e^+e^-$ or $q\overline{q}$ pairs are far smaller,
$\sim 10^{-10}$. The latest upper bounds on 
the WIMP-nucleon scalar cross section from CDMS
\cite{Akerib} are in the range
$10^{-42.4}-10^{-41}~{\rm cm^2}\approx 10^{-15}-10^{-13.6}
~\gev^{-2}$, compared to $\sigv\approx 8.6\times 10^{-9}
~\gev^{-2}(x_f\sqrt{\gstar}/15\gstars)$ for s-wave annihilation
and $\sigma_0\approx 1.7\times 10^{-7}~\gev^{-2}
(x_f^2\sqrt{\gstar}/150\gstars)$ for p-wave annihilation,
so very low branching ratios for $mm\to({\rm standard~model~pairs})$
may be reasonable, if not expected. 

Astronomical observations could hardly have missed the production of pair fireballs with energy $\sim M_{ann}$ on timescales $\sim GM_{BH}$.  Moreover, a blast wave composed of $e^\pm$ pairs, photons and
baryons with a total energy that is even a modest 
fraction of $\Mann$ (say $0.001-0.01$) could deposit enough 
energy to substantially alter if not completely eject the
interstellar medium of a galaxy.  One possibility is that the branching ratios are $\lesssim 10^{-10}$ for $mm\to e^+e^-$ or $q{\overline{q}}$ and $\lesssim 10^{-5}$ for $mm\to\nu{\overline{\nu}}$, as discussed above.  Alternatively, either black holes do not form from annihilating dark matter, or else the dark matter is in some other form that does not annihilate, such as axions.

\begin{acknowledgments}

R. A. Vanderveld has been supported by an NSF Graduate Research Fellowship and a Cornell NASA Space Grant Fellowship.

\end{acknowledgments}

\end{document}